\documentclass[aps,prl,twocolumn]{revtex4}
\usepackage{bm}
\usepackage{amssymb}
\usepackage{graphicx}
\usepackage{epstopdf}
\usepackage{amsmath}

\newcommand{\BEQ}{\begin{equation}}
\newcommand{\EEQ}{\end{equation}}
\newcommand{\BEA}{\begin{eqnarray}}
\newcommand{\EEA}{\end{eqnarray}}

\newcommand{\p}{\partial}
\newcommand{\s}{\sigma}
\renewcommand{\t}{\tau}

\newcommand{\nn}{\nonumber }

\makeatletter
\newcommand\figcaption{\def\@captype{figure}\caption}
\makeatother

\begin{document}

\title{ General phase-diagram of multimodal ordered and disordered
  lasers in closed and open cavities} \author{F. Antenucci$^{1,2}$,
  C.Conti$^{1,3}$, A. Crisanti$^{1,3}$ and L. Leuzzi$^{2,1}$}
\email{luca.leuzzi@cnr.it} \affiliation{$^1$ Dipartimento di Fisica,
  Universit\`a di Roma ``Sapienza,'' Piazzale A. Moro 2, I-00185,
  Roma, Italy\\ $^2$ IPCF-CNR, UOS {\em Kerberos} Roma, Piazzale
  A. Moro 2, I-00185, Roma, Italy \\ $^3$ ISC-CNR, UOS {\em Sapienza},
  Piazzale A. Moro 2, I-00185, Roma, Italy }

\begin{abstract}
We present a unified approach to the theory of multimodal laser
cavities including a variable amount of structural disorder.  
  A general mean-field theory is studied for waves in media with variable
non-linearity and randomness.  Phase diagrams are reported in terms of
optical power, degree of disorder and degree of non-linearity, 
tuning between closed and open cavity scenario's.
In the thermodynamic limit of infinitely many modes  the theory predicts four
distinct regimes:
a continuous wave behavior for low power, a standard mode-locking
laser regime for high power and weak disorder, 
a random laser for high pumped power and
large disorder and an intermediate regime of phase locking
occurring in presence of disorder below the lasing threshold.  
\end{abstract}

\maketitle

In describing cavity-less lasers with randomly placed scatterers,
generically called random lasers (RLs)
\cite{Lethokov68,Lawandy94,Cao99,Sebbah02,Cao05,Tureci06,Tureci08,Wiersma08,Leuzzi09,Turitsyn10,Zaitsev10}, the most challenging issue is the interplay between
disorder and non-linearity.  In RLs the
lasing is due to stimulated amplification of light spatially localized
in leaky stochastic resonators \cite{Lethokov68}. If disorder is dominant and non-linearity
negligible, stimulated amplification of light can be hindered
because of diffusion.  Conversely, if the
structural disorder is weak, the effect on nonlinear
evolution is marginal and does not modify the standard laser features. 
Competition occurs when wave scattering affects the degree of localization and non-linearity couples the
localized modes.  
Various recent experimental results
show that the coupling between modes changes the degree of spatial
localization and their spatial and spectral correlations
\cite{Leonetti13b,Leonetti13c,Leonetti13d}.

In this Letter, we report a theoretical analysis comprehensively describing all the possible regimes,
accounting for the fact that light modes exhibit a distribution of localization
lengths, and an interaction determined by the overall energy. 
In a general statistical mechanical
framework, we predict specific transitions from incoherent to coherent
regimes, both  in the case of standard mode-locking lasers in a closed
cavity and for cavity-less disordered systems with strong gain. The latter
displaying a {\em glassy} coherent behavior. We also define and consider, as an independent parameter, a
{\em degree of non-linearity}.  As {\em glassy} we mean that (i) a
sub-set of modes out of an extensive ensemble of localized passive
modes are activated in a non-deterministic way 
\footnote{We stress that by non-deterministic we do not mean
  chaotic, since chaos is, actually, deterministic.  Chaos is an apart
  dynamic phenomenon occurring in laser systems, \cite{Weiss91} but it
  does not, actually, affect the presence or absence of glassiness, as, e.g., shown in Ref. [\onlinecite{Crisanti96}]. 
  Chaos
  is not a necessary feature for RLs, nor it is sufficient one. } and (ii) the
whole set of activated modes behaves cooperatively and belongs to one state out of many possible ones.
The system properties can be represented by a corrugated
landscape composed of many valleys separated by high mountains and
hidden passes. The overall coherence arises from the trapping in a metastable state in the landscape.

We identify four different equilibrium phases: continuous wave
(CW$\sim$), phase-locked wave (PLW$\natural$), standard ML laser
(SML$\parallel$), random lasing (RL$\star$).  In previous work, phases of modes were retained as the only relevant dynamic
variables \cite{Leuzzi09,Conti11}. Here we remove this {\em {quenched
    amplitude}} approximation and provide a general
picture of all the regimes attainable in a multi-modal
laser at any degree of pumping, disorder, and cavity leakage.

\noindent {\em The complex amplitude model. ---} For a closed cavity,
localized modes form a complete set and the electro-magnetic field
$\bm{E}(\bm r, t)$ can be expanded in terms of normal modes
$\bm{E}_n({\bm r})$ with time-dependent complex amplitudes $a_n(t)$\cite{Conti11}.
In open cavities a continuous spectrum of radiation modes is also present.  
The contributions of radiative and localized modes can be separated
by a suitable projection onto two orthogonal subspaces
\cite{Hackenbroich03,Viviescas03}. This leads to an effective theory
on the subspace of localized modes in which they
exchange a linear off-diagonal effective damping coupling.  Radiation
losses and gain are accounted for by additional linear terms (diagonal
when the net gain is homogeneous), and the presence of a thermal bath
is represented by the fluctuations due to the spontaneous emission.  Nonlinear
couplings arise from gain saturation and from the optical Kerr effect.
At equilibrium with the pump mechanism, the complex amplitudes
$a_n(t)$ are linked by a constraint given by the total optical
intensity inside the system ${\cal E} = \epsilon N = \sum_{m=1}^N
|a_m|^2$, where $N$ is the number of modes and $\epsilon$ the average energy per mode.
The general Hamiltonian, derived by different approaches
\cite{Angelani06long,Leuzzi09,Conti11}, reads
\BEA
{\cal H}=-\Re\Biggl[&&\hspace*{-3mm}
\frac{1}{2}\sum^{1,N}_{n_1, n_2} J_{\vec{n}_2}  a_{n_1} a_{n_2}^* 
\label{f:H}
\\
\nonumber
&&\hspace*{-.3cm}+\frac{1}{4!}\hspace*{-.1cm}
\sum^{n_k=1,N}_{\omega_{n_1}+\omega_{n_3}=\omega_{n_2}+\omega_{n_4}}
\hspace*{-.8cm}
J_{\vec{n}_4}  a_{n_1} a_{n_2}^* a_{n_3}  a_{n_4}^*\Biggr]
\nonumber
\EEA
\noindent where $J_{\vec{n}_p}=J_{n_1\ldots n_p}$ and the second sum
ranges over all distinct $4$-ples for which the so-called ML condition
holds:
%
$\omega_{n_1}-\omega_{n_2}
+\omega_{n_3}-\omega_{n_4}=0$.
The  coupling coefficient $J_{\vec{n}_{4}}$ represents the spatial
overlap of the electromagnetic fields modulated by
non-linear $\chi^{(3)}(\{\omega\};\bm r)$ susceptibility: 
\begin{eqnarray}
 J_{\vec{n}_4}&=&\frac{\imath}{2}\prod_{j=1}^4 \sqrt{\omega_{n_j}}
\label{f:J}
\int_V d^3r~
\chi^{(3)}_{\vec{\alpha}_4}(\{\omega_{\vec{n}_4}\};
\mathbf r) 
\\
&& \hspace*{2.5cm}
\times ~ E^{\alpha_1}_{n_1}(\mathbf r) E_{n_2}^{\alpha_2}(\mathbf r) E_{n_3}^{\alpha_3}(\mathbf r)
 E_{n_4}^{\alpha_4}(\mathbf r)
\nonumber
 \end{eqnarray} 
with $\alpha_j = x,y,z$, and $\vec{n}_4=\{n_1, n_2, n_3, n_4\}$.  The
linear coefficient $J_{\vec{n}_{2}}$ yields different contributions depending
on medium randomness and cavity leakage:
\begin{eqnarray}
J_{\vec{n}_2}&=& J_{n_1} \delta_{n_1n_2}+ J^{\rm rad}_{\vec{n}_2}+ J^{\rm inh}_{\vec{n}_2}
\label{f:J2}
\\
 J^{\rm inh}_{\vec{n}_2}&=&\frac{\imath}{2} \sqrt{\omega_{n_1}\omega_{n_2}}
\int_V d^3r~
\chi^{(1)}_{\vec{\alpha}_2}(\mathbf r) 
 ~ E^{\alpha_1}_{n_1}(\mathbf r) E_{n_2}^{\alpha_2}(\mathbf r)
\label{f:J2_inh}
\end{eqnarray}
  The linear diagonal terms of $J_{\vec{n}_{2}}$ depend on gain and
  loss profiles for the passive modes.  A possible non-uniform
  distribution of the gain - and the related inhomogeneous linear
  susceptibility $\bm \chi^{(1)}(\bm r)$ - yields the spatial
  overlap of localized eigenmodes, i. e., $J^{\rm inh}$, Eq. (\ref{f:J2_inh}).
    Besides, in the
  {\em open cavity} scenario, the linear off-diagonal coupling terms
  also account for the presence of a continuous spectrum, and they
  correspond to the effective damping contribution $J^{\rm rad}$
  obtained integrating out radiation modes
  \cite{Hackenbroich03,Viviescas03}.  
  Taking the purely nonlinear
  interaction of a discrete set of modes corresponds to the {\em
    strong cavity limit}, $J^{\rm rad}=0$, with
  homogeneous gain (because of orthogonality of
  $\{\bm E_n\}$'s it is $J^{\rm inh}=0$).  Only linear diagonal terms remain in this limit.

We build a mean-field theory in which the system is fully connected,
that is, the network of interacting modes is a complete graph.  This
amounts to adopt a {\em narrow bandwidth} approximation for the gain
profile in which $\omega_n\simeq \omega_0$, for each
$n=1,\ldots,N$ \footnote{ More specifically,
  $|\omega_j-\omega_k|<\delta\omega$, for each $j,k=1,\ldots,N$, where
  $\delta\omega$ is the line-width of the intensity spectrum.  Indeed,
  in most of RLs, it is not necessary that the resonant ML
  condition for having four modes interact is satisfied exactly
  \cite{MeystreBook}.}.  
This is the case in the so-called {\em dispersive} RLs with
very low finesse and a sensitive narrowing of the bandwidth above
threshold, in which many modes oscillate in a relative small bandwidth
and are so densely packed in frequency that their linewidths overlap.
Consistently with this approximation we take a constant effective net
gain profile in the bandwidth: $J_{n}= g(\omega_n)\simeq
g(\omega_0)=g_0$, implying $\sum_{n_1,n_2} J_{\vec{n}_2}a_{n_1}a_{n_2}
= g_0 {\cal E} + \sum_{n_1 \neq n_2} J_{\vec{n}_2}a_{n_1}a_{n_2}$.

The open cavity model in the narrow bandwidth approximation can be viewed as an extension of the so-called
  spherical $2$+$p$ model
  \cite{Nieuwenhuizen95,Crisanti04c,Crisanti06,Crisanti07,Crisanti13}
 yielding a far reacher variety of physical scenarios.

The off-diagonal linear terms and the non-linear terms may, in
general, be disordered because modes 
display different degree and shape of localizations \cite{Conti08PhC,Fallert09}. 
The constituents of the integrals in Eq. (\ref{f:J},\ref{f:J2_inh}) are very difficult to calculate
from first principles. The only specific form of the
non-linear susceptibility has been computed by Lamb
\cite{Lamb64,LambBook} for few-modes standard lasers and no
analogue study for RLs has been performed so far, to our knowledge.   
The overlap integrals in a disordered system can
be regarded as a sum over many random variables.
Correspondingly, the probability distribution of the  couplings $J_{\vec{n}_p}$ can be assumed 
to be 
 Gaussian:
\BEA
P(J_{\vec{n}_p})=
\sqrt{\frac{N^{p-1}}{2\pi J^2_p}}
\exp\left\{
-\frac{N^{p-1}}{2J_p^2}
\left[J_{\vec{n}_p}-\frac{J_0^{(p)}}{N^{p-1}}\right]^2
\right\}
\label{def:Gauss}
\EEA with $p=2,4$.  To simplify the computation and its presentation
we will take real-valued interaction couplings.  This amounts, e. g., to
neglect the effect of group velocity in the diagonal linear part and
the Kerr lens effect in the nonlinear term, but does not change the
generality of the qualitative picture.  This
is the most general Hamiltonian model for laser systems that one can
consider. Indeed, also adding further non-linear terms
($J_{\vec{n}_p}$ with $p=3,5,6,\ldots$) does not alter the qualitative
behavior at the transition from continuous wave to (standard or random)
lasing regimes.

\noindent {\em The external parameters ---} In order to yield a comprehensive description, we introduce the {\em
  degrees of non-linearity} $\alpha_0,\alpha$ - varying in the
interval $[0,1]$ - and suitable interaction energy scales $J_0,J$, for
the ordered and disordered component, respectively: 
\BEA
J_0^{(4)}=\alpha_0 J_0; & \alpha_0
=\left[\frac{J_0^{(2)}}{J_0^{(4)}}+1\right]^{-1}\!\!\!\!\!\!; &
J_0=J_0^{(2)}+J_0^{(4)} \\ J_4=\alpha J;\quad & \alpha
=\left[\frac{J_2}{J_4}+1\right]^{-1} ; &\,\, J=J_2+J_4 \EEA
\noindent The {\em degree of disorder} of a given system with coupling parameter scales $J,J_0$
is, then, defined as
$R_J = J/J_0$.


The average energy per mode $\epsilon$ is related to the so-called
{\em pumping rate} ${\cal P}$ induced by the pumping laser source in
the RL, or proportional to the optical power in the cavity
for the standard laser.  In the present work it is defined as ${\cal
  P}\equiv \epsilon \sqrt{J_0/k_B T}=\epsilon\sqrt{\beta J_0} $ where
$T$ is the heat-bath temperature. 
It encodes the experimental evidence
that decreasing temperature \cite{Wiersma01} or increasing the total
power \cite{Leonetti10} yields qualitatively similar behaviors.
  The factor $J_0$ is a material dependent parameter function of the angular
  frequency $\omega_0$ of the peak of the average spectrum,
  cf. Eq. (\ref{f:J}), and it is volume independent.  
  
To summarize, the  parameters of interest are:\\
\begin{tabular}{ll}
\vspace*{-.3cm}\\
Optical power per mode&$\epsilon$\\
Heat-bath thermal energy &$k_B T=1/\beta$\\ 
Cumulative coupling average & $J_0=J_0^{(2)}+J_0^{(4)}$\\
Cumulative mean square disp. & $J=J_2+J_4$\\
\vspace*{-.3cm}\\
Pumping rate& ${\cal P}=\epsilon\sqrt{\beta J_0} $ \\
Disorder degree& $R_J=J/J_0$\\
Non-linearity degree (ordered) & $\alpha_0=J_0^{(4)}/J_0$ \\
Non-linearity degree (disordered) & $\alpha=J_4/J$. \\
\end{tabular}

\vskip .1 cm \noindent We will consider here $\alpha=\alpha_0$
for simplicity, but cases with different degrees of non-linearity in
ordered and disordered contributions can be also analyzed.

\noindent {\em Statistical mechanics with replicas. ---}
We study the model by means of the replica trick \cite{Edwards75}.
This enables to calculate the average free energy $f$ in the one step Replica Symmetry Breaking (1RSB) Ansatz as a function of generalized order parameters,
as detailed in the Appendix.
We find a  series of order parameters
 describing the physical regimes: (i) the intensity coherence of
 activated modes $m$; (ii) the phase coherence $r_d$; (iii-iv) the
 overlap parameters $q_{0,1}$ and (v) the RSB parameter $x$. The
 latter three specify degree and kind of glassiness.
The free energy reads  reads
\BEA
&&2 \beta f(q_0,q_1,r_d,m)= 2\beta f_0+
(1-x)w(q_1,q_1)
\nonumber
\\
\nonumber
&&\qquad +x~w(q_0,q_0) -w(r_d,1)- 2 k(m)
-\ln(1-r_d) 
\\
&&\qquad  -\ln {\cal X}_1
-\frac{1}{x}\ln\frac{{\cal X}_0}{{\cal X}_1}
-\frac{2q_0-m^2}{{\cal X}_0}
\label{f:f}
\EEA
with
\BEA
&&{\cal X}_1\equiv 1+r_d-2 q_1 \, ; \quad 
{\cal X}_0\equiv {\cal X}_1+ 2x(q_1-q_0)
\\
&&w(t,u)\equiv \xi_2(t^2+u^2)+\frac{\xi_4}{2}(t^4+u^4+4t^2u^2)\qquad
\label{def:w}
\\ &&k(m)\equiv k_2 |m|^2+k_4 |m|^4
\\ &&
\label{f:xi24}
\xi_2=\frac{\beta^2\epsilon^2}{4}J_2^2 \, ; \qquad
\xi_4=\frac{\beta^2\epsilon^4}{6}J_4^2 
\\ 
\label{f:k24}&&k_2=\frac{\beta
  \epsilon}{4}J_0^{(2)}\, ; \qquad k_4=\frac{\beta
  \epsilon^2}{96}J_0^{(4)} 
\EEA 
The self-consistency saddle point
equations for the order parameters $m$, $r_d$, $q_{0,1}$, $x$ are
given in the Appendix.

 \noindent These parameters determine all relevant thermodynamic phases
(refer to Fig. \ref{fig:PhDi_P_R_a}):

\begin{figure}[t!]
\center \includegraphics[width=1.\columnwidth]{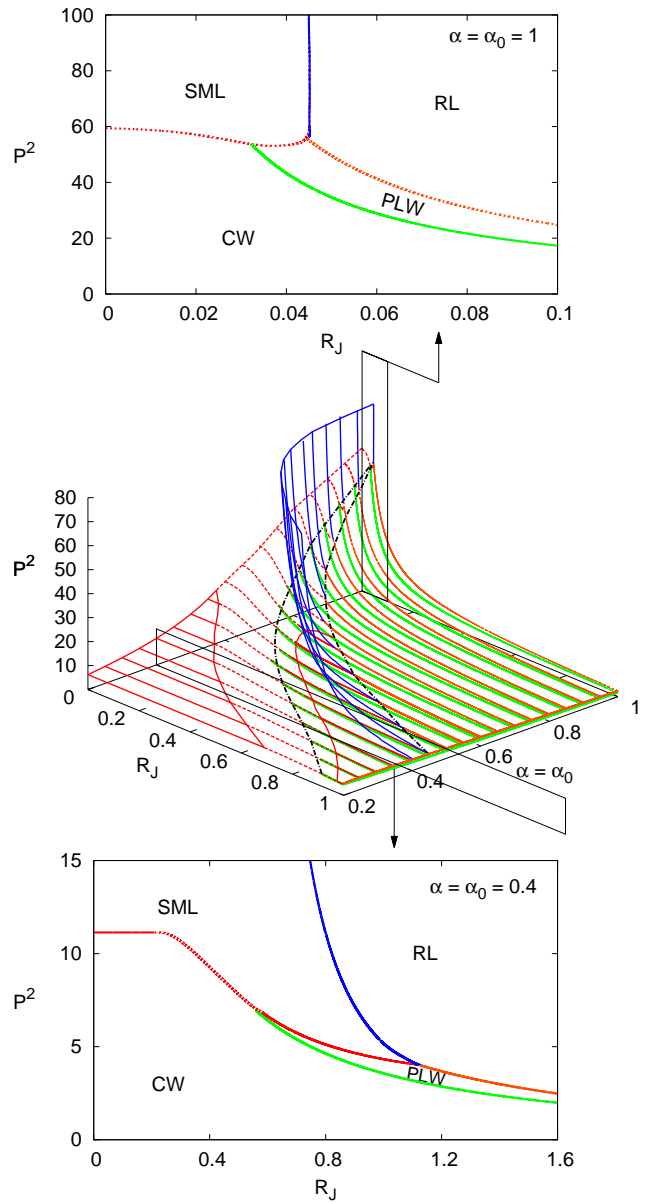}
\vspace*{-6mm}
\caption{Phase diagrams in the (${\cal P}^2, R_J, \alpha=\alpha_0$)
  space.  For low disorder only the SML and CW
  phases occur varying pumping and degree of linearity.  As disorder
  increases the intermediate PLW phase arise between CW and SML. For
  strong disorder RL replaces SML above the pumping threshold line.
  For any $R_J$, for low $\alpha_0$ the transition driven by ${\cal
    P}$ is continuous in the order parameters, whereas for high
  non-linearity ($\alpha>\alpha_{\rm nl}=0.6297$) it is discontinuous. Insets: phase diagrams in the
  (${\cal P}^2, R_J$) plane for closed (right: $\alpha=\alpha_0=1$ and
  open ($\alpha=\alpha_0=0.4$) cavity. }
\label{fig:PhDi_P_R_a}
\end{figure}

$\sim\qquad $ {\em Continuous wave regime} (CW). $\qquad$ For small optical
power, all modes oscillate
independently in a CW incoherent noisy regime and the energy
equally fluctuates among all passive modes.  At low ${\cal P}$, for
any degree of quenched disorder $R_J$ and non-linearity $\alpha$,
all parameters are $q_{01}=m=r_d=0$ ($x$ is irrelevant when $q_0=q_1$).

$\natural\qquad$ {\em Phase locking wave regime} (PLW). $\qquad$ For
non-zero disorder, increasing ${\cal P}$ the system undergoes a
transition to a thermodynamic phase in which the mode phases 
lock on one given value, without stimulated
amplification.  Considering the complex amplitudes as continuous
spherical spins in two dimensions, this corresponds to all spins
pointing in the same direction though their intensity is freely
oscillating.  This phase has no counter-part in statistical mechanical
models studied so far.  The phase coherence
parameter $r_d$ is non-zero, whereas $q_{0,1}=m=0$.  For
increasing non-linearity $\alpha$, PLW occurs at lower and lower
pumping rate ${\cal P}$.
 
$\parallel\qquad${\em Standard Mode-locking laser} (SML). $\qquad$ For
large ${\cal P}$ and $R_J=0$ a localization transition occurs and
the intensity is shared by  activated modes, all of them
oscillating coherently. This corresponds to standard
passive ML laser systems \cite{Haus00}, where a passive transition in
${\cal P}$ is predicted as a paramagnetic/ferromagnetic transition in
Ref.  [\onlinecite{Gordon03}] 
\footnote{The $R_J=0$ limit  of our theory confirms previous estimates.
To quantitatively compare we have rescaled our
  parameters in the notation of Ref.  [\onlinecite{Gordon03}], where
  the transition point was given in the parameter $T/\gamma_s P_0^2 =
  8/{\cal P}^2 $, with $P_0 = \epsilon$ and $ \gamma_s = J_0/8$. In
  the quenched amplitude approximation we have $8/{\cal P}^2= 0.1828$
  and in the free amplitude model we find $8/{\cal P}^2 = 0.1357$,
  both compatible with previous results.}.  
We further find that the CW/SML transition takes place also 
in presence of a limited amount of disorder $R_J\gtrsim 0$.
The magnitude $m\neq 0$ in this regime so that light modes 
are coherent in intensity and stimulated amplification occurs.

$\star\qquad${\em Random lasing} (RL). $ \qquad$ For high pumping rate and
strong disorder, the tendency to oscillate synchronously is
frustrated, resulting in a glassy phase representing the random laser
regime.  Modes are all coherent in phase ($r_d\neq 0$) but not in
intensity ($m=0$), acquiring fixed random values.  The
thermodynamic phase is glassy with one step of RSB: $q_1>q_0=0$ and
$x\neq 0$
\footnote{This is strictly true in the whole glassy light phase above
  threshold when non-linearity is dominating and the transition is
  discontinuous in the order parameters, see {Appendix}.
In systems where $\alpha$ is small, instead, a glassy phase is still
present but it is of a different nature and, actually, described by an
infinite step RSB. Since, though, this ``linearly runned" transition
is continuous in the overlap, rather than discontinuous as for
$\alpha>\alpha_{\rm nl}$, the 1RSB is a fully justified approximation next
to the lasing where $ q_1  \gtrsim q_0 \gtrsim 0 $.}.

 \noindent {\em Phase Diagram. ---}
In Fig. \ref{fig:PhDi_P_R_a} we show the ${\cal P}^2$, $\alpha$, $R_J$
diagram in the main 3D plot. In the top inset the closed cavity projection 
for $\alpha=1$ is displayed and in the bottom inset  an open cavity  instance, $\alpha=0.4$.

For low degree of disorder $R_J$ there is a  threshold between a CW (or
PLW) phase and a SML; for large $R_J$ the threshold is to a RL.  
The CW/SML threshold line is plotted as solid (continuous transition) or
dotted (discontinuous transition) dark (red) line and it occurs for $R_J\gtrsim 0$. 
The locus of tricritical (SML,CW,PWL) transition points
is plotted as a black dashed-dotted line.  

As $R_J$ is still small but
a bit larger than the tricritical point, increasing the pumping one first has a CW/PWL transition
denoted by a continuous light gray (green) line and then a
PWL/SML. The latter can be continuous (solid dark-gray/red line) or
discontinuous (dotted). 

  For larger $R_J$ we exclusively observe
 PWL/RL transition.  
The nature of the PWL/RL transition depends on the degree of non-linearity; it occurs
 both continuously, for $\alpha<\alpha_{\rm nl} $
or discontinuously, for $\alpha>\alpha_{\rm nl}$, see Appendix. 
In Fig. \ref{fig:PhDi_P_R_a} it is $\alpha_{\rm nl}= 0.6297$. 

Supposing, further,
that in an optically active material the degree of disorder $R_J$
could be continuously  changed, the phase diagram also predicts a
transition between SML and RL for pumping above threshold, represented
by the nearly vertical solid (blue) transition lines.  The locus of
the tricritical points RL/SML/PWL is plotted as a second black
dashed-dotted line in the main plot.
\begin{figure}
\includegraphics[width=1\columnwidth]{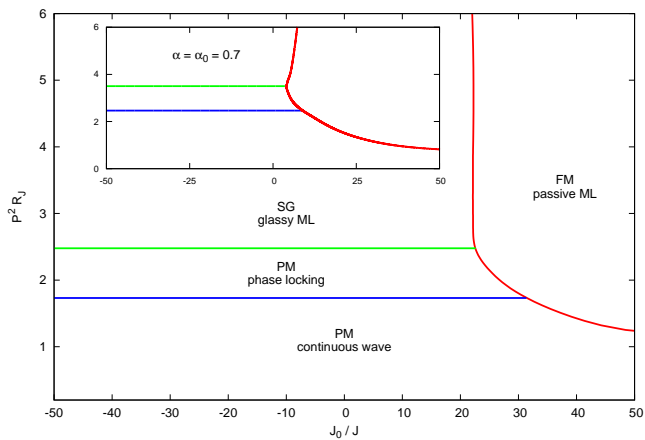}
\caption{Phase diagram in ${\cal P}^2 R_J=\beta \epsilon^2 J$
  vs. $J_0/J=1/R_J$ for closed cavity $\alpha=\alpha_0=1$ for $J_0/J$ ranging
  from negative to positive. Inset: open cavity with $\alpha=0.7$.}
\label{fig:PhDi_P2_J0}
\end{figure}

\noindent {\em Mode locking without saturable absorber ---} A key point in
the present study is that the transition from continuous wave to
standard passive mode-locking (CW$\to$ SML) only occurs for a strictly
positive value of the coupling coefficient $J_0$, as shown in Figs.
(\ref{fig:PhDi_P_R_a},\ref{fig:PhDi_P2_J0}).  This formally
corresponds to the presence of a saturable absorber in the cavity
\cite{Conti11}.  In RLs such a device is not present, and,
hence, the occurrence of this lasing transition is not to be given for
granted. However, in Fig. (\ref{fig:PhDi_P_R_a}) it is shown that
starting from a standard laser supporting passive mode-locking and
increasing the disorder, the CW/SML transition acquires the character
of a glassy mode-locking CW/RL transition. This is present also for
$J_0<0$, as explicitly shown in Fig. (\ref{fig:PhDi_P2_J0}).  It is
far from trivial that the latter mode-locking transition, ruled out
for ordered lasers without a saturable absorber ($J_0 \ll J$),
spontaneously occurs as an effect of the quenched disorder, i.e., for large
enough $R_J$. The physical origin is the mode coupling due to the open cavity configuration.

\noindent {\em Conclusions ---} We developed a theory for light wave
systems in optically active random media with narrow bandwidth and in
which all activated mode localization spatially overlap with each
other.  The theory is based on the most general Hamiltonian for an
open system, including non-linearity and coupling to a thermal bath
that measures the amount of energy transferred to the system by an
external pumping system at equilibrium.  This corresponds to a
mean-field fully connected disordered system in statistical mechanics.
We derive the most general phase diagram, ranging from closed
cavities that correspond to a standard laser, to completely open
cavities, representing the random lasers.  The resulting
picture, given in terms of order parameters that represent the
correlations among amplitudes, phases, and their cross correlation,
furnishes a number of different equilibrium phases affected by non-linearity and disorder.  
These include standard passive-mode locking in standard lasers, and different
coherent regimes attainable in random lasers.  The reported
results open the way to further investigations, as the study of open
quantum systems, testing theoretical models for the statistical mechanics of disordered
systems, employing variable coherence sources for applications in
spectroscopy and microscopy, and developing novel techniques for mode locking and ultra-short pulse generation not requiring saturable absorbers.

\noindent {\em Acknowledgments ---} The research leading to these results has
 received funding from the People Programme (Marie Curie Actions) of
 the European Union's Seventh Framework Programme FP7/2007-2013/ under
 REA grant agreement n¡ 290038, NETADIS project, from the European Research
Council through ERC grant agreement no. 247328 - CriPheRaSy project - and from the Italian
 MIUR under the Basic Research Investigation Fund FIRB2008 program,
 grant No. RBFR08M3P4, and under the PRIN2010 program, grant code
 2010HXAW77-008.

\appendix 
\section*{Appendix: Replica Calculations}
\label{App}
In this appendix  we carry out explicit replica calculations first  in the general framework and
then  in the specific one step Replica Symmetry Breaking (RSB) Ansatz. In the following, we show the complete derivation of the free energy functional and the self-consistency equations for the relevant order parameters describing the regimes reported in the paper.

In our {\em narrow bandwidth} approximation, given a line-width $\delta
\omega$ any frequency difference $|\omega_i-\omega_j|\ll \delta
\omega$.  In other words the band-width $\Delta \omega$ of the
spectrum of the optically active medium is comparable with, or less
than, the line-width of each mode and the system finesse is $f\equiv
\Delta\omega/\delta\omega<1$.  

The diagonal part of the pairwise interaction of Eq. (3), with real
$J$'s, represents the net gain frequency profile

\BEQ
\nn
J_{n}=g(\omega_n)\ .
\EEQ
 In this narrow-band approximation the
corresponding contribution to the Hamiltonian becomes:
\begin{equation}
-\frac{1}{2}\sum_{n} J_{n}|a_n|^2 = -\frac{1}{2} g(\omega_0) \epsilon N
\label{diag}
\end{equation}
that is, an irrelevant constant in the dynamics.

Naming real and imaginary part of the complex amplitudes as 
\BEQ
a=\s+\imath \t
\nn
\EEQ
and neglecting the diagonal term, Eq. (\ref{diag}), the Hamiltonian  -
Eq. \ref{f:H} of the main text -  becomes
\BEA
{\cal H}&=&-\sum_{n_1<n_2}J_{n_1n_2}(\s_{\vec{n}_2}+\t_{\vec{n}_2})\\
\nonumber
&&
-\sum_{n_1<n_2<n_3<n_4}J_{n_1 n_2 n_3 n_4}(\s_{\vec{n}_4}+\t_{\vec{n}_4}+\Theta_{\vec{n}_4})
\EEA
with
\BEA 
 \Theta_{\vec{n}_4} &\equiv& \frac{1}{3}\left(
\psi_{n_1n_2,n_3n_4}+
\psi_{n_1n_3,n_2n_4}+
\psi_{n_1n_4,n_2n_3}
\right) \, ,
\nn
\\
 \psi_{n_1n_2,n_3n_4}&\equiv& \s_{n_1n_2}\t_{n_3n_4}+\s_{n_3n_4}\t_{n_1n_2} \, ,
\nn
\\
\s_{n_x n_y}&=& \s_{n_x}\s_{n_y}\qquad;\hspace*{1cm} 
\s_{\vec{n}_p}\,\,=\,\,\,\prod_{j=1}^p \s_{n_j} \; .
\nn
\EEA

In a nutshell, we study the model thermodynamics by means of the replica trick \cite{Edwards75}:
\begin{eqnarray}
&&\langle \ln Z[J]\rangle_{P(J)}= \lim_{n\to 0}\frac{1}{n}\left[ \langle Z^n[J]\rangle_{P(J)}-1\right]
\nonumber
\\
&&Z^n[J]=\hspace{-.5cm}\sum_{\{\{a^{(1)}\},\ldots,\{a^{(n)}\}\}}\hspace*{-.5cm}
\exp\left\{ -\beta \sum_{\nu=1}^n
{\cal H}[\{a^{(\nu)}\}|\{J\}] \right\}
\nonumber
\end{eqnarray}
where $\{a^{(\nu)}\}=\s^\nu+\imath \t^\nu$ is the mode amplitude configuration in the replica $\nu$.
The random interaction network is copied $n$ times, the
partition function is computed for $n$ copies and, eventually, the
analytic continuation to real $n$ is computed in the $n\to 0$ limit,
yielding the expression for the average free energy
\BEQ
\beta f({\cal Q}) = -\frac{1}{N}\langle \ln Z[J]\rangle_{P(J)} \bigr|_{S.P.}
\label{f:frep}
\EEQ
 as a function of a given saddle point solution for the set of order
 parameters ${\cal{Q}}$.

Rescaling  the degrees of freedom as 
\BEQ
\sigma \to \sqrt{\frac{\epsilon}{2}} ~ \sigma \quad ; \quad \t \to \sqrt{\frac{\epsilon}{2}} ~\t \ ,
\nn \EEQ
the average over
the $J$'s distribution  of the   partition function of the spherical model replicated $n$ times turns out to be
\BEA \langle{Z^n[J]}\rangle_{P(J)}&\!=\!&\!\!\int_{{\cal S}_n} \!\!\!
{\cal D}\s~ {\cal D}\t~
\exp \Biggl\{ \frac{2k_2}{N}\sum_{n_1<n_2}\sum_{a=1}^n
\left(\s_{\vec{n}_2}^a+\t_{\vec{n}_2}^a\right)
\nn
 \\ && 
 \nn + \frac{ \xi_2}{2 N}\sum_{n_1<n_2}\sum_{ab}
\left(\s_{\vec{n}_2}^a+\t_{\vec{n}_2}^a\right)
\left(\s_{\vec{n}_2}^b+\t_{\vec{n}_2}^b\right) \\ &&
 \nn 
+ \frac{24
  k_4}{ N^3}\sum_{n_1<n_2<n_3<n_4}\sum_{a=1}^n
\left(\s^a_{\vec{n}_4}+\t^a_{\vec{n}_4}+\Theta^a_{\vec{n}_4}\right)
\\ \nn && +\frac{9 \xi_4}{4 N^3}\sum_{n_1<n_2<n_3<n_4}\sum_{ab}
\left(\s^a_{\vec{n}_4}+\t^a_{\vec{n}_4}+\Theta^a_{\vec{n}_4}\right)
\\ && 
\nn \quad \qquad \qquad \qquad \qquad \times
\left(\s^b_{\vec{n}_4}+\t^b_{\vec{n}_4}+\Theta^b_{\vec{n}_4}\right)
\Biggr\}
\\
\label{Z} \EEA
with
\begin{equation}
\nonumber
{\cal  D}\s=\prod_{i=1}^N\prod_{a=1}^{n}d\s_i^{(a)}   \, , 
\end{equation}
\noindent and where the integral is over the $n$ hyper-spheres ${\cal S}_N$ given by the spherical constraint
\BEQ
\sum_{i=1}^N\left(\s_i^{(a)}+\t_i^{(a)}\right)^2 = 2N.  
\label{spher} \EEQ
The average pump energy per mode $\epsilon$ now enters only in 
the parameters $\xi_2$, $\xi_4$, $k_2$ and $k_4$ as defined in Eqs. (\ref{f:xi24}-\ref{f:k24}) of the main text. 

 Introducing as in the following the overlap parameters between replicas 

\BEA Q_{ab}&=&\frac{1}{2N}\sum_{j=1}^N\left( \s^a_j\s^b_j+\t^a_j\t^b_j
\right) 
\label{Qdef}
\\ R_{ab}&=&\frac{1}{2N}\sum_{j=1}^N\left(
\s^a_j\s^b_j-\t^a_j\t^b_j \right)
\label{Rdef}
\\ T_{ab}&=&\frac{1}{N}\sum_{j=1}^N\s^a_j\t^b_j
\EEA
and the
 magnetization vectors in the replica space (the super-index $R$ stands for real, $I$ stand for imaginary)
 \BEA
 m^R_a&=&\frac{1}{N}\sum_{j=1}^N\s^a_j
\label{mRdef}
\\ m^I_a&=&\frac{1}{N}\sum_{j=1}^N\t^a_j  \ , 
\label{mIdef}
\EEA 
with further manipulations we sum in the partition function 
over the configurations of real and imaginary parts of the complex amplitudes,
 ending up with the replicated free energy
and the saddle point equations for the order parameters. 

 Before
reporting the calculation, though, we observe that, for symmetry
reasons, in absence of an external field linearly coupled to $\s$ or
to $\t$ or, more generally, coupled to any simple function of the phase
of the complex amplitude $\phi=\arctan (\t/\s)$, it holds
\begin{equation}
T_{ab}=T_{ba}=0; \qquad   \forall a,b \qquad \ .
\label{T_0}
\end{equation}

Generalizing a well known procedure for $p$-spin models
\cite{Gardner85,Crisanti92}, the above defined parameters
Eqs. (\ref{Qdef}), (\ref{Rdef}), (\ref{mRdef}) and (\ref{mIdef}) are
inserted in Eq. (\ref{Z}) as Dirac deltas and these are expressed in their
Fourier form, introducing further parameters, the Lagrange multipliers
$\lambda_{ab}$, $\mu_{ab}$, $\kappa^R_a$, $\kappa^I_a$

\noindent
Carrying out this approach  we end up with 
\BEA
\langle{Z^n[J]}\rangle_{P(J)}=\int {\cal D}\bm Q ~{\cal D}\bm \lambda
\exp\Bigl\{
-N G(\bm Q,\bm \lambda)\Bigr\}
\EEA
with $\bm Q=\{Q, R, \bm m\}$ and $\bm \lambda=\{\lambda,
\mu,  \bm \kappa\}$, and  where we introduce the $2n$ column vectors 
 \BEA
 \nn
 \bm m &\equiv& \left(
m_1^R, \ldots, m_n^R, m_1^I, \ldots, m_n^I\right)^T
\\
\nn 
\bm \kappa &\equiv& \left(
\kappa_1^R, \ldots, \kappa_n^R, \kappa_1^I, \ldots, \kappa_n^I\right)^T 
\EEA
to shorten the notation.
The replicated free energy functional, then, reads
\BEA
G(\bm Q,\bm \lambda) &\equiv& A(\bm Q) + B(\bm Q,\bm \lambda)-\ln {\cal Z}(\bm \lambda)
\label{G_Q_lambda}
\\
A(\bm Q) &\equiv& -\frac{\xi_2}{2}\sum_{ab} \left(Q_{ab}^2+R_{ab}^2\right)
\label{A_Q}
\\
&&
-\frac{\xi_4}{4}\sum_{ab}\left(Q_{ab}^4+R_{ab}^4+4Q_{ab}^2R_{ab}^2\right)
\nn
\\
&&
-k_2 ~ \bm m^T\cdot \bm m
- k_4 ~\left[\bm m^T\cdot \bm m\right]^2
\nn
\\
B(\bm Q,\bm\lambda)&=&\sum_{ab}\left(\lambda_{ab}Q_{ab}+\mu_{ab}R_{ab}\right)
+\bm \kappa^T \cdot \bm m
\label{B_Q_lambda}
\\
{\cal Z}(\bm \lambda)&=&
\int {\cal D}\s \, {\cal D}\t \, 
\exp\Bigl\{\sum_a\bigl(
\kappa_a^R\s_a+\kappa_a^I\t_a
\bigr)
\\
\nn
&&
+\frac{1}{2}\sum_{ab}\bigl[
\s_a(\lambda_{ab}+\mu_{ab})\s_b
+ \t_a(\lambda_{ab}-\mu_{ab})\t_b
\bigr]  
\Bigr\}
\\
&=&\frac{(2\pi)^n}{\sqrt{\det(-\lambda-\mu)\det(-\lambda+\mu)}}
\nn \\
\nn
&&\qquad\qquad\times
\exp\Bigl\{
-\frac{1}{2}
(\bm\kappa^R)^T(\lambda+\mu)^{-1}\bm\kappa^R
\nn
\\
\nn
&&\qquad\qquad\qquad \,\,\quad-\frac{1}{2}
(\bm\kappa^I)^T(\lambda-\mu)^{-1}\bm\kappa^I
\Bigr\}
\EEA
The latter term in Eq. (\ref{G_Q_lambda}) can be written as 
\BEA
-\ln {\cal Z}(\bm\lambda)&=&\frac{1}{2}\ln \det W-\frac{1}{2}
\bm\kappa^TW^{-1}\bm\kappa
\label{lnZrep}
\\
W&\equiv& -\left(\begin{array}{cc} \lambda+\mu & 0\\
0& \lambda-\mu
\end{array}
\right)
\EEA

Saddle point equations ${\p G}/{\p \bm\lambda}=0$ obtained deriving with respect to 
$\lambda_{ab}$, $\mu_{ab}$, and $\kappa_a$, turn out to be, respectively 
\BEA
&&2 Q-\bm m^R\bm m^R-\bm m^I\bm m^I  = -(\lambda+\mu)^{-1}-(\lambda-\mu)^{-1}
\nn
\\
&&
2 R-\bm m^R\bm m^R +\bm m^I\bm m^I= -(\lambda+\mu)^{-1}+(\lambda-\mu)^{-1}
\nn
\\
&&
\bm m =  W^{-1}  \bm \kappa
\nn
\EEA
 yielding the self-consistency relations
\BEA
\label{SP_lambda1}
&&\lambda+\mu=-\left(Q+R-\bm m^R\bm m^R\right)^{-1}
\\
\label{SP_lambda2}
&&\lambda-\mu=-\left(Q-R-\bm m^I\bm m^I\right)^{-1}
\\
\label{SP_lambda3}
&&\bm \kappa^{R,I}=\left(Q+R-\bm m^{R,I}\bm m^{R,I}\right)^{-1} \bm m^{R,I}
\EEA

Using Eqs. (\ref{SP_lambda1}-\ref{SP_lambda3}) and  observing that 
of Eq. (\ref{B_Q_lambda}) plus the last term in Eq. (\ref{lnZrep}) sum up to zero, that is,
\BEA
\nn
&&B(\bm Q,\bm \lambda)-\frac{1}{2}
\bm\kappa^TW^{-1}\bm\kappa  
\\
\nn
&&\qquad\qquad =\sum_{ab}\left(\lambda_{ab}Q_{ab}+\mu_{ab}R_{ab}\right)
+\frac{1}{2}\bm \kappa^T \cdot \bm m=0 \ , 
\EEA

The subsequent saddle point equations from stationarity of
 $G(\bm Q)$ with respect to the elements of  $Q,R$ and $\bm m$ are
 \BEA
&& 2 \xi_2 Q_{ab} +  2 \xi_4 \left(Q_{ab}^3+2Q_{ab}R_{ab}^2\right) 
\label{SP_Q1}
\\
&&\qquad = 
- (Q+R-\bm m^R\bm m^R)_{ab}^{-1} - (Q-R-\bm m^I \bm m^I)_{ab}^{-1}
\nn
\\
&& 2 \xi_2  R_{ab} + 2 \xi_4 \left(R_{ab}^3+2R_{ab}Q_{ab}^2\right) 
\label{SP_Q2}
\\
&&\qquad = 
- (Q+R-\bm m^R\bm m^R)_{ab}^{-1} + (Q-R-\bm m^I \bm m^I)_{ab}^{-1}
 \nn
\\
&& 2 k_2  m^R_{a} + 4 k_4 \left[(m^R_{a})^3+m^R_{a}(m^I_{a})^2\right]  
\label{SP_Q3}
\\
&&\qquad\qquad\qquad=\left[\bm m^R (Q+R -\bm m^R\bm m^R)^{-1}\right]_a
\nn
\\
&& 2 k_2 m^I_{a} + 4 k_4 \left[(m^I_{a})^3+m^I_{a}(m^IR_{a})^2\right]
\label{SP_Q4}
\\
&&\qquad \qquad \qquad=\left[\bm m^I (Q-R -\bm m^I\bm m^I)^{-1}\right]_a
\nn
 \EEA

 \noindent
Using Eqs. (\ref{SP_Q1}-\ref{SP_Q4}), Eqs. (\ref{SP_lambda1}-\ref{SP_lambda3}) lead to
\BEA
\lambda_{ab} &=&  \xi_2  Q_{ab} +   \xi_4  Q_{ab}\left(Q_{ab}^2+2R_{ab}^2\right)
\\
\mu_{ab}&=&  \xi_2   R_{ab} +   \xi_4   R_{ab}\left(R_{ab}^2+2Q_{ab}^2\right)
\\
\bm \kappa &=& 2 k_2  \bm m + 4 k_4  \bm m (\bm m^T\cdot \bm m)
\EEA

\noindent
Eventually,
using that
\BEA
&& \ln\det(Q \pm R-\bm m^{R,I} \bm m^{R,I}) =
\label{G_Q}
\\
&=& \ln\det(Q \pm R)
\nn
- (\bm m^{R,I})^T(Q \pm R)^{-1}\bm m^{R,I}  + O(n^2)
\EEA

\noindent
we write the free energy Eq. (\ref{G_Q_lambda}) in the form
\BEA
G(\bm Q)&=&-\frac{1}{2}\sum_{ab} w(Q_{ab},R_{ab})-\sum_a k(\bm m_a)
\\
&&
-\frac{1}{2}\ln\det(Q+R)
-\frac{1}{2}\ln\det(Q-R)
\nn
 \\
\nn
&&
  -n \ln \frac{ \epsilon}{2} +\frac{1}{2}(\bm m^R)^T (Q+R)^{-1}\bm m^R
\\
\nn
&&+\frac{1}{2}(\bm m^I)^T(Q-R)^{-1}\bm m^I
\EEA

\noindent
with, cf. Eqs. (10-11) of the main text,
\BEA
w(t,u)&\equiv& \xi_2 (t^2+u^2)+\frac{\xi_4}{2}(t^4+u^4+4t^2u^2)
\nn
\\
k(\bm m)&\equiv& k_2 ~ \bm m^T\cdot \bm m + k_4 (\bm m^T\cdot \bm m)^2  \, .
\nn
\EEA
The diagonal parts will now be $Q_{aa}=1$, because of the complex spherical constraint, cf. Eqs. (\ref{spher},\ref{Qdef}),
and we will, further, term $R_{aa}=r_d$.

The stationarity equations take the form 
\BEA
\label{SP_QR1}
&&\Lambda(Q_{ab}, R_{ab})+ |\bm m|^2 B(\bm m) =
\\
\nn 
&&\qquad\qquad=-(Q+R)^{-1}_{ab}-(Q-R)^{-1}_{ab}
\\
\label{SP_QR2}
&&\Lambda(R_{ab}, Q_{ab})+ \bigl[(m^R)^2-(m^I)^2\bigr] B(\bm m)
\\
\nn
&&\qquad\qquad = -(Q+R)^{-1}_{ab}+(Q-R)^{-1}_{ab}
\\
\label{SP_QR3}
&&m^{R,I} B(\bm m)= m^{R,I} \sum_b (Q\pm R)^{-1}_{ab}
\EEA
where 
\BEA
\Lambda(t,u)& \equiv& 2t [\xi_2+\xi_4(t^2+2u^2)]\nn
\\
B(\bm m) &\equiv & 2 k_2 + 4 k_4 |\bm m|^2
\nn
\EEA
We have dropped the replica dependence of $\bm m^T=(m^R,m^I)$ since
observables depending on single replica index do not break replica
symmetry \cite{Crisanti04d} and the sum along each row/column of any function of the $\bm
Q$ matrices is invariant under column/row permutation.  

We now notice
that any solution to the above equations (\ref{SP_QR3}) implies
\BEQ
m^R m^I=0
\EEQ
and this is a consequence of Eq. (\ref{T_0}). Choosing, then, without loss of generality, $m^I=0$,
Eqs. (\ref{SP_QR1}-\ref{SP_QR3}) are solved by 
\BEQ
R_{ab}=Q_{ab} \qquad a\neq b
\EEQ
For the diagonal part still holds $Q_{aa}=1$, $R_{aa}=r_d$.

\begin{figure}[t!]
\vspace*{1cm}
\includegraphics[width=1.\columnwidth]{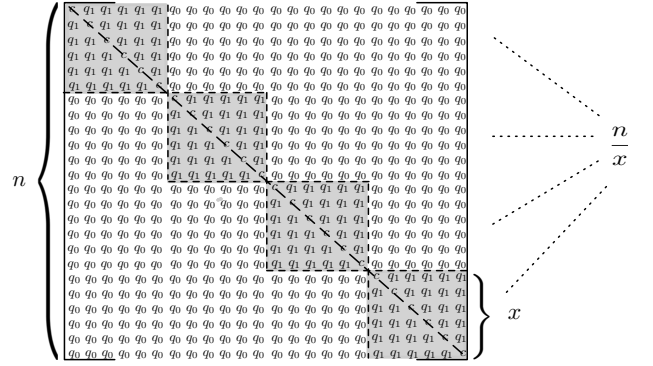}
\caption{1RSB scheme for the elements of the $Q_{ab}$ overlap parameter matrix. As $n\to 0$, 
$x$ becomes a continuous variable in the intervall $[0:1]$. The elements of $Q_{ab}$ acquire two values: $q_0$, if in an off-diagonal block, and $q_1>q_0$ inside the diagonal blocks (diagonal excluded).}
\label{fig:1RSB_scheme}
\end{figure}

{\em Replica Symmetry Breaking $\qquad$---$\qquad$}
We will not show here the stability analysis of the various Replica
Symmetry Breaking Ans{\"a}tze involved in given regions of the phase
diagram. Even though the most general stable solution to the problem
is within a Full RSB scheme \cite{Parisi80}, that is a {\em continuous} overlap function
$q(x)$ in the interval $x\in[0,1]$, for the sake of clarity in the presentation, and without loss of generality for what concerns the study of the threshold behaviors,
we will adopt the one step
RSB scheme, cf. Fig. \ref{fig:1RSB_scheme}. Calling, as in the figure, $x$ the number of elements in a row of a diagonal block of $Q_{ab}$, 
the elements take two values according to the rule
\BEA
Q_{ab}=\Biggl\{\begin{array}{cc} q_1 & \mbox{if $N(a/x)$=$N(b/x)$ } \\
\vspace*{-.2cm}
\\
q_0 & \mbox{if $N(a/x)\neq N(b/x)$}
\end{array}
\EEA
being $N(a/x)$ the integer part of $a/x$.

In this Ans{a}tz the free energy take the form shown in Eq. (\ref{f:f}) of the main text, 
where the self-consistency saddle point equations for the order parameters $m$, $r_d$, $q_{0,1}$, $x$ are given by
 \BEA
m~B(m) &=& \frac{m}{\cal X}_0 
\label{eq:B}
\\ \Lambda(q_0,q_0)+m^2 B^2(m)&=&\frac{2
  q_0}{{\cal X}_0}
\\ \Lambda(q_1,q_1)-\Lambda(q_0,q_0)&=&\frac{2(q_1-q_0)}{{\cal
    X}_0{\cal X}_1}
\\ 
\Lambda(r_d,1)-\Lambda(q_1,q_1)&=&\frac{1}{1-r_d}-\frac{1}{{\cal
    X}_1} \\ w(q_1,q_1)-w(q_0,q_0)&=&\frac{1}{x^2}\ln \frac{{\cal
    X}_0}{{\cal X}_1} -\frac{2(q_1-q_0)}{x {\cal X}_0} 
    \\ \nonumber
&&+\frac{2(q_1-q_0)(2q_0-m^2)}{{\cal X}_0^2} \EEA 

where the last equation corresponds to the stationarity in the $x$ parameter \footnote{We name  the integer number of elements in a diagonal block in Fig. \ref{fig:1RSB_scheme} and its analytic continuation in the $n\to 0$ limit invariably by $x$.}.
These are the parameters that describe all relevant thermodynamic phases involved,
 as displayed in Fig. \ref{fig:PhDi_P_R_a} of the main text.

For $\alpha>\alpha_{\rm nl}\simeq (3-1.76382 \epsilon) (3-1.03703 \epsilon^2)^{-1}  $ ($\simeq 0.6297$ for $\epsilon =1$), that is both for open and closed
cavity, the 1RSB solution is stable above pumping threshold and it
reduces discontinuously to an RS solution below threshold.  This is the exact solution. For small
$\alpha<\alpha_{\rm nl}$, that is, for extremely open cavities, instead, the actual
solution is Full RSB and it reduces continuously to a Replica Symmetric solution
below threshold ($q_1-q_0\to 0$).  In virtue of this continuity at the
transition, the 1RSB solution very well qualitatively and quantitatively approximates
the Full RSB solution next to the PWL/RL transition line.

We postpone to a more technical paper the complete description of the complex amplitude model with fixed optical power in terms of replica symmetry breaking
solutions and their stability analysis
\footnote{For what concerns the purely real $2+4$ spherical spin model
  the reader can refer to
  [\onlinecite{Crisanti04c,Crisanti06,Crisanti13}].}.


\end{document}